# $\mathcal{PT}$ symmetry in optics beyond the paraxial approximation

Changming Huang,[1] Fangwei Ye,[1,*] Yaroslav V. Kartashov,[2] Boris A. Malomed,[3] and Xianfeng Chen[1]

[1]*State Key Laboratory of Advanced Optical Communication Systems and Networks, Department of Physics and Astronomy, Shanghai Jiao Tong University, Shanghai 200240, China*
[2]*Institute of Spectroscopy, Russian Academy of Sciences, Troitsk, Moscow Region 142190, Russia*
[3]*Department of Physical Electronics, School of Electrical Engineering, Faculty of Engineering, Tel Aviv University, Tel Aviv 69978, Israel*
*Corresponding author:fangweiye@sjtu.edu.cn*

The concept of the $\mathcal{PT}$ symmetry, originating from the quantum field theory, has been intensively investigated in optics, stimulated by the similarity between the Schrödinger equation and the paraxial wave equation governing the propagation of light in guiding structures. We go beyond the paraxial approximation and demonstrate, solving the full set of the Maxwell's equations for the light propagation in deeply subwavelength waveguides and periodic lattices with balanced gain and loss, that the $\mathcal{PT}$ symmetry may stay unbroken in this setting. Moreover, the $\mathcal{PT}$ symmetry in subwavelength guiding structures may be *restored* after being initially broken upon the increase of gain and loss. Critical gain/loss levels, at which the breakup and subsequent restoration of the $\mathcal{PT}$ symmetry occur, strongly depend on the scale of the structure.© 2014 Optical Society of America

The canonical quantum theory postulates that every physical observable is associated with a Hermitian operator, guaranteeing that eigenvalues of this operator are real. Nevertheless, systems with non-Hermitian Hamiltonians, which obey the parity-time ($\mathcal{PT}$) symmetry [1], may also have purely real spectra. The $\mathcal{PT}$-symmetry implies that the Hamiltonian includes a complex potential obeying condition $V(x)=V^*(-x)$. A transition to a complex spectrum, which is called the $\mathcal{PT}$-symmetry breaking, occurs with the increase of the strength of the imaginary part of the potential [1].

The $\mathcal{PT}$-symmetry has been a subject of active investigations in optics, as photonic structures offer a real platform for the implementation of this concept [2-5]. The above-mentioned condition, $V(x)=V^*(-x)$, is translated into the requirement of the spatial symmetry of the refractive index and antisymmetry of the gain/loss profile in guiding optical structures. The concept of the $\mathcal{PT}$ symmetry was extended to nonlinear couplers [6,7], periodic lattices [8-12], and nonlinear complex potentials [13,14]. Remarkable effects, such as unidirectional invisibility [15] and the absence of reflection [16] were demonstrated in $\mathcal{PT}$-symmetric optical structures.

Previous works on the $\mathcal{PT}$ symmetry in optics assumed paraxial propagation in shallow refractive-index landscapes, with all spatial scales being much larger than the wavelength of light. Accordingly, the light transmission is governed by the Schrödinger equation for the amplitude of the electromagnetic field, implementing the similarity to $\mathcal{PT}$-symmetric Hamiltonians in the quantum theory.

The advance in the nanofabrication technologies suggests the use of nano-scale structures, where the light propagation is directly governed by the Maxwell's equations (ME). Numerous works have addressed the propagation of light in diverse passive subwavelength guiding settings [17-20]. However, the $\mathcal{PT}$ symmetry has never been studied in deeply subwavelength structures with the balanced gain and loss. The stark difference between the scalar paraxial Schrödinger equation and the full ME system for the vectorial fields suggest the following questions: Can the concept of the $\mathcal{PT}$ symmetry, that was originally formulated in terms of the Schrödinger equation be carried over into the ME realm? And then, how do vectorial and nonparaxial effects, inherently present in the ME, affect the $\mathcal{PT}$ symmetry? It is relevant to add that the $\mathcal{PT}$ symmetry was recently applied to metamaterials [16,21,22] (still, relying on the paraxial description). The use of the ME should be essential in this context too, as such materials are built of subwavelength elements.

In this Letter, we address two basic active subwavelength structures: a single waveguide, and a periodic lattice, with the real and imaginary parts of the dielectric permittivity being, respectively, even and odd functions of the transverse coordinate. By solving the full ME system, we demonstrate that the $\mathcal{PT}$ symmetry *persists* in these settings at the subwavelength scale. In contrast to the paraxial approximation, where the $\mathcal{PT}$ symmetry is always broken above a critical value of the gain/loss coefficient, in the nonparaxial regime this symmetry may be *restored* after its initial breakup, following the increase of the gain/loss strength.

We consider the propagation of a TM-polarized light beams (i.e., only $E_x, E_z, H_y$ components of the electric and magnetic fields are nonzero) along the $z$ axis in a medium whose dielectric permittivity is modulated in the trans-

verse direction, $x$. The evolution of the field components is governed by the reduced ME system:

$$i\frac{\partial E_x}{\partial z} = -\frac{1}{\varepsilon_0 \omega}\frac{\partial}{\partial x}\left(\frac{1}{\varepsilon_{rel}}\frac{\partial H_y}{\partial x}\right) - \mu_0 \omega H_y,$$
$$i\frac{\partial H_y}{\partial z} = -\varepsilon_0 \varepsilon_{rel} \omega E_x, \quad (1)$$

with $E_z = (i/\varepsilon_0 \varepsilon_{rel} \omega)\partial H_y/\partial x$. Here $\varepsilon_0$ and $\mu_0$ are the vacuum permittivity and permeability, $\omega$ is the field frequency, and $\varepsilon_{rel}(x) = \varepsilon_{bg} + \varepsilon^{re}(x) + i\varepsilon^{im}(x)$ the relative permittivity of the $\mathcal{PT}$-symmetric structure, with background permittivity $\varepsilon_{bg}$. In what follows below, we fix wavelength $\lambda = 632.8$ nm, $\varepsilon_{bg} = 2.25$, and consider the following permittivity landscapes:

$$\varepsilon_{rel}(x) = \varepsilon_{bg} + p\,\text{sech}^2(x/d) + i\alpha\,\text{sech}(x/d)\tanh(x/d), \quad (2a)$$

$$\varepsilon_{rel}(x) = \varepsilon_{bg} + p\cos(2\pi x/d) + i\alpha\sin(2\pi x/d), \quad (2b)$$

for an isolated waveguide and periodic lattice, respectively, where $p$ and $\alpha$ determine the modulation depths of the real and imaginary parts of the dielectric permittivity, while $d$ is the waveguide's width or lattice period. Varying $d$, one can pass from paraxial ($d \gg \lambda$) to subwavelength ($d < \lambda$) regime. Even and odd refractive-index and gain-loss profiles in Eq. (2), $n^{re}(x)$ and $n^{im}(x)$, correspond to $\varepsilon_{rel}(x) \equiv [n^{re}(x) + in^{im}(x)]^2$.

Guided modes in the single waveguide [Eq.(2a)] are looked for as $[E_x(x,z), H_y(x,z)] = [E_x(x), H_y(x)]\exp(ibz)$, where complex propagation constant $b = b_r + ib_i$ identifies three possible types of solutions: (i) bound modes with $b_r > \varepsilon_{bg}^{1/2}$ and $b_i = 0$, propagating without attenuation or growth, (ii) decaying or growing bound modes with $b_r > \varepsilon_{bg}^{1/2}$ and $b_i \neq 0$, and (iii) delocalized modes with $b_r < \varepsilon_{bg}^{1/2}$. Following the common definition, the $\mathcal{PT}$ symmetry remains unbroken as long as the system's spectrum remains purely real for a given set of $p, \alpha, d$ values.

Eigenmodes of the waveguide were found using an eigen-system package (Fortran LAPACK), and checked by a commercial mode-solver[24]. Figure 1 shows dependencies $b_{r,i}(\alpha)$ for different widths of the waveguide. For relatively broad (but subwavelength) waveguides with $d = 120$ nm, the picture in Fig. 1(a) is similar to that known in the paraxial regime. Although the permittivity is complex, for relatively small $\alpha$ the propagation constants of all the modes remain *purely real*, i.e., the $\mathcal{PT}$-symmetry holds. When $\alpha$ approaches the so-called exceptional value, $\alpha_{ex} \approx 1.95$ in the present setting, a pair of modes with complex-conjugate propagation constants emerge. The magnitude of $|b_i|$ increases with $\alpha$ without any signature of saturation.

The results displayed in Fig. 1 are produced by solving the full vectorial ME in the subwavelength structure, therefore the persistence of the conventional features of the $\mathcal{PT}$ symmetry, *viz.*, purely real spectra under the complex permittivity, and its spontaneous breaking at the exceptional point, should not be taken for granted, as these properties were previously produced solely by the Schrödinger equation.

Another noteworthy property of the $\mathcal{PT}$-symmetry is revealed when the size of the waveguide is downscaled to the deep-subwavelength regime [$d = 60$ nm in Fig. 1(b)]. In this case, the system keeps a real spectrum at $\alpha < \alpha_{ex}$. The distributions of $|E_x|$ and $|E_z|$ in the corresponding fundamental mode are symmetric [Fig. 2(a)]. In this deeply nonparaxial regime, the amplitude of the longitudinal field is comparable to that of the transverse ones. The spectrum becomes complex at $\alpha > \alpha_{ex}$, when a pair of complex-conjugate eigenvalues appears, and the distribution of $|E_{x,z}|$ becomes notably asymmetric – depending on the sign of $b_i$, the mode concentrates mostly in the region with the gain or loss [Fig. 2(b)].

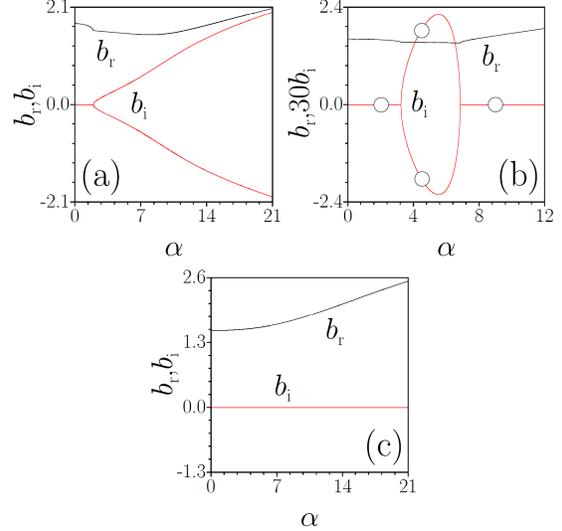

Fig. 1. (Color online) Real $b_r$ and imaginary $b_i$ parts of the propagation constant versus $\alpha$ for the single waveguide with $d = 120$ nm (a), $d = 60$ nm (b), and $d = 30$ nm (c). Circles in (b) correspond to the eigenmodes shown in Fig. 2, and examples of the propagation dynamics in Fig. 3. In all the cases, $p = 1.7$.

A striking property of the nonparaxial $\mathcal{PT}$-symmetry is that further increase of the gain/loss strength results in the *restoration* of the real spectrum, at $\alpha > \alpha_{rest}$. Then, the spectrum remains purely real even for very large values of $\alpha$. This is accompanied by the restoration of the symmetry of the $|E_{x,z}|$ profiles, which become very narrow [Fig. 2(c)]. The restoration of the $\mathcal{PT}$-symmetry occurs only in sufficiently narrow waveguides, i.e., it is a truly nonparaxial effect. Moreover, further decrease of the waveguide's width to $d = 30$ nm results in complete *elimination* of the $\mathcal{PT}$-symmetry-breaking effect, with the spectrum remaining purely real for *any* $\alpha$ [Fig. 1(c)].

An opposite scenario was recently demonstrated in Ref. [23]: broken $\mathcal{PT}$ symmetry was recovered and broken again in a model with two gain-loss coefficients.

Typical mode evolution scenarios, for different values of $\alpha$ and $d = 60$ nm, are presented in Fig. 3. The modes below the exceptional point propagate as stationary ones [Fig. 3(a)], while the modes in the region of the broken $\mathcal{PT}$ symmetry, with complex-conjugate values of $b$, are either amplified [Fig. 3(b), $b_i < 0$] or attenuated [Fig. 3(c), $b_i > 0$]. At $\alpha > \alpha_{rest}$, the modes with the restored symmetry again exhibit the stationary propagation [Fig. 3(d)].

Figures 4(a) and 4(b) summarize properties of the propagation spectra in the $(d, \alpha)$ plane for $p = 0.3$ and

$p = 1.7$, respectively. In addition to the $\mathcal{PT}$-symmetry-restoration regions (shown in red), the figures reveal the dependence of the exceptional point $\alpha = \alpha_\text{ex}$, at which the initial symmetry breaking occurs, on the waveguide's width $d$. For relatively wide waveguides with $d > 300$ nm, the symmetry breaking occurs at $\alpha_\text{ex} \approx p$, which is nearly independent of $d$. However, as $d$ decreases, $\alpha_\text{ex}$ grows and diverges at some point (at $d \approx 44$ nm for $p = 0.3$, and at $d \approx 53$ nm for $p = 1.7$), indicating that the $\mathcal{PT}$ symmetry is always preserved in the sufficiently narrow waveguide. Localized modes do not exist in blue domains of the $(d, \alpha)$ plane.

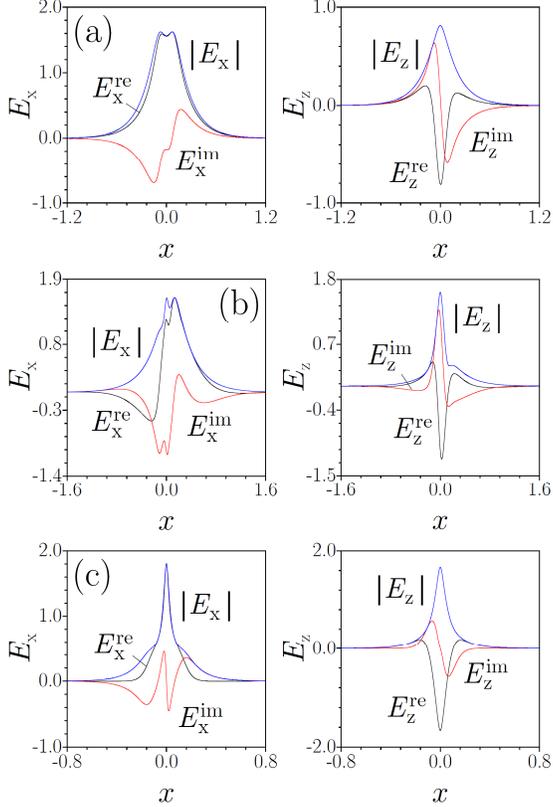

Fig. 2. (Color online) Profiles of the guided modes at (a) $\alpha = 2.0$, (b) $\alpha = 4.5$, and (c) $\alpha = 9.5$. Two modes at $\alpha = 4.5$ are mutually conjugate, therefore only the growing one is shown. Fields $E_{x,z}$ are plotted in dimensionless units, while transverse coordinate $x$ is measured in $\mu$m. In all the cases, $p = 1.7$.

In subwavelength periodic lattices with symmetric $\varepsilon^\text{re}(x)$ and antisymmetric $\varepsilon^\text{im}(x)$ permittivity profiles [Eq.2(b)], the eigenmodes are extended Bloch waves, whose profiles can be found as $H_y^{(n)}(x, z) = \tilde{H}_y^{(n)}(x) \exp[ikx + ib^{(n)}z]$, where $\tilde{H}_y^{(n)}(x)$ is a $d$–periodic function, $k$ is the Bloch momentum, and $n$ is the band index, with similar expressions for field components $E_{x,z}$ are similar. Dependencies of propagation constant $b^{(n)} = b_r^{(n)} + ib_i^{(n)}$, in different bands, on momentum $kd/\pi$ are strongly affected by the value of $\alpha$.

For relatively large periods, $d \sim \lambda$, two upper allowed bands, $b_r^{(1)}$ and $b_r^{(2)}$, that were separated by the gap at $\alpha = 0$, approach each other with the increase of $\alpha$, until they merge at points $kd/\pi = \pm 1$ at $\alpha_\text{ex} = p$. In the merger region, we have $b_r^{(1)} = b_r^{(2)}$ and $b_i^{(1)} = -b_i^{(2)} \neq 0$. The subsequent increase of $\alpha$ leads to the merger of higher-order Bloch bands, so that the $\mathcal{PT}$-symmetry remains broken at $\alpha_\text{ex} > p$. The situation changes drastically for small lattice periods, as shown in Fig. 5 for $d = 230$ nm and

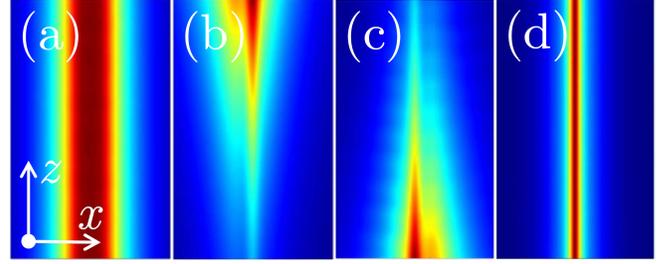

Fig. 3. (Color online) The propagation of the eigenmodes in the 60 nm-wide waveguide with $p = 1.7$ at (a) $\alpha = 2.0$, (b,c) $\alpha = 4.5$ (both growing and decaying modes are shown), and (d) $\alpha = 9.5$. The propagation distance is 40 $\mu$m in (a),(d) and 2 $\mu$m in (b),(c).

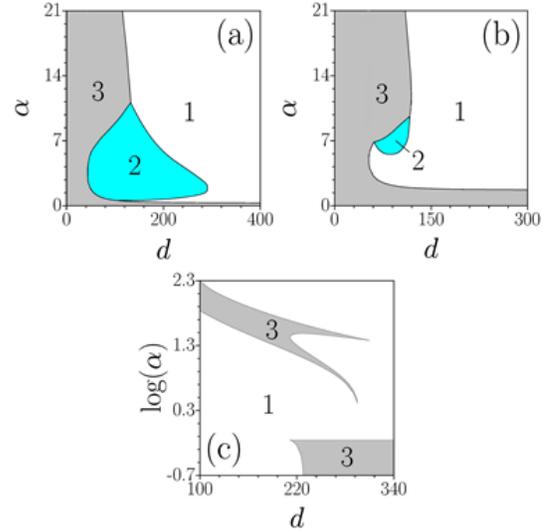

Fig. 4. (Color online) Existence domains for eigenmodes of an isolated waveguide (a),(b), and Bloch modes of the periodic lattice (c), on the $(d, \alpha)$ plane. Region 1 corresponds to broken $\mathcal{PT}$ symmetry, in the region 2 there are no localized modes, region 3 corresponds to the unbroken or *restored* $\mathcal{PT}$ symmetry. In all the cases, $d$ is plotted in nm. In (a) $p = 0.3$, in (b) $p = 1.7$, and in (c) $p = 0.7$.

$p = 0.7$. In this case, one encounters two bands separated by a gap at $\alpha = 0$ [Fig. 5(a)], which exhibit the merger at $\alpha > p$ [Fig. 5(b)]. However, at sufficiently large strengths of the imaginary part of the periodic potential, $\alpha > 12.6$, the first gap *reopens*, and the $\mathcal{PT}$ symmetry is *completely restored*, with a purely real spectrum of the eigenmodes [Fig. 5(c)]. The first gap remains open within a finite interval of $12.6 < \alpha < 16.3$, and then shrinks again [see Fig. 6(a) which display $b_r^{(n)}(\alpha)$ dependencies corresponding to the bottom of the first band, top of the second band, and top of the third band], so that the system *re-enters* the broken-$\mathcal{PT}$-symmetry state. Then, restoration of the $\mathcal{PT}$-symmetry occurs again at even higher values of $\alpha$ [see an open gap between $b_r^{(1)}, b_r^{(2)}, b_r^{(3)}$ curves in Fig. 6(b) corresponding to a purely real spectrum]. The domain where the $\mathcal{PT}$-symmetry experiences the restoration in the subwavelength lattice is shown red in Fig. 4(c).

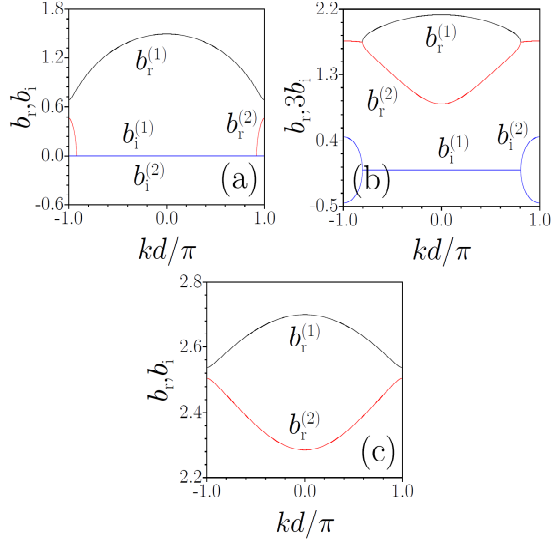
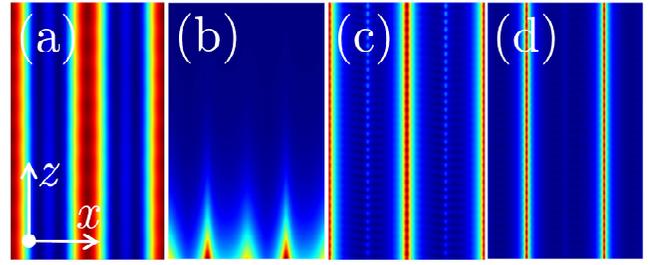

Fig. 7. (Color online) The propagation dynamics of Bloch waves with $k=\pi/d$ in the array with period $230$ nm and $p=0.7$. (a) A wave from the first band at $\alpha=0.6$, (b) a decaying wave at $\alpha=6.0$ from the region where the first and second bands fuse, (c) a wave from the first band at $\alpha=14.2$, (d) a wave from the second band at $\alpha=14.2$. The propagation distance is $20$ $\mu$m in (a),(c),(d), and $2$ $\mu$m in (b).

Fig. 5. (Color online) Dependencies $b_r^{(n)}(k)$ and $b_i^{(n)}(k)$ for the first and second allowed bands of the lattice with $d=230$ nm and $p=0.7$, at $\alpha=0.6$ (a), $\alpha=6.0$ (b), and $\alpha=14.2$ (c). In panel (c), zero imaginary parts, $b_i^{(1)}, b_i^{(2)}=0$, are not shown.

Propagation scenarios for Bloch waves corresponding to the cases depicted in Fig. 5 are shown in Fig. 7. Below the $\mathcal{PT}$-symmetry-breaking point the Bloch modes corresponding to $k=\pi/d$ propagate steadily [Fig. 7(a)], as the spectrum is real. Above the critical point, the Bloch modes with $k=\pi/d$ grow or decay [Fig. 7(b)]. When the $\mathcal{PT}$-symmetry is restored, one again observes stationary propagation [Figs. 7(c),(d)].

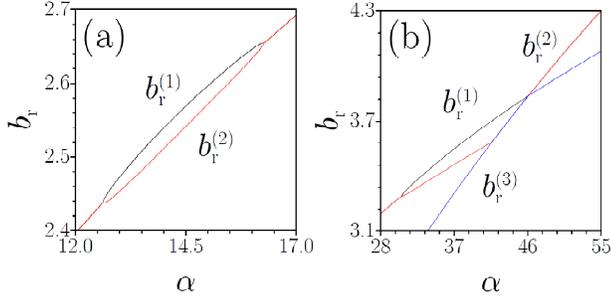

Fig. 6. (Color online) Dependencies $b_r^{(n)}(\alpha)$ in the lattice with $d=230$ nm, $p=0.7$ in the intervals of $\alpha$ where reopening of the first finite gap occurs.

To summarize, we have extended the concept of the $\mathcal{PT}$-symmetry in optical media to the setting governed by the full ME system. Subwavelength $\mathcal{PT}$-symmetric guiding structures may support full sets of eigenmodes with purely real spectra, i.e., the $\mathcal{PT}$-symmetry is preserved, provided that the strength of the gain/losses is not too large. In contrast to paraxial systems, where the $\mathcal{PT}$-symmetry remains broken above the exceptional point, subwavelength structures allow restoration of the symmetry beyond this point. The $\mathcal{PT}$ symmetry may stay intact (*endless*) at all values of the gain/losses, if the transverse scale of the guiding structure is small enough. The present work is focused on TM modes, leaving similar analysis for TE waves for subdequent work.